\documentclass[fleqn,twoside]{article}

\usepackage{espcrc2}
\usepackage[psamsfonts]{amsfonts}

\newcommand{\tr}{\mathop{\rm tr}\nolimits}
\newcommand{\Tr}{\mathop{\rm Tr}\nolimits}
\newcommand{\SU}{\mathop{\rm SU}}
\newcommand{\U}{\mathop{\rm {}U}}
\newcommand{\rmd}{{\rm d}}
\newcommand{\Gammait}{{\mit\Gamma}}

\title{Axial anomaly in the reduced model: Higher representations%
      \thanks{Talk presented by T. Inagaki at Lattice 2003.}}
\author{Teruaki Inagaki\address{Graduate School of Science and Engineering,
Ibaraki University, Mito 310-8512, Japan},
Yoshio Kikukawa\address{Department of Physics, Nagoya University, Nagoya
464-8602, Japan} and
Hiroshi Suzuki\address{Department of Mathematical Sciences, Ibaraki University,
Mito 310-8512, Japan}}

\begin{document}
\begin{abstract}
The topological charge in the $\U(N)$ vector-like reduced model can be defined
by using the overlap Dirac operator. We obtain its large $N$ limit for a
fermion in a general gauge-group representation under a certain restriction of
gauge field configurations which is termed $\U(1)$ embedding.
\vspace{1pc}
\end{abstract}

\maketitle

\section{Introduction}
It had been unclear how to define the topological charge in the reduced model
which is given by a zero-volume limit of lattice gauge theory. In
ref.~\cite{Kiskis:2002gr}, the authors proposed to define the topological
charge in a ``fermionic way'' by using the overlap Dirac operator and showed
that there actually exist non-trivial topological sectors. In
ref.~\cite{Kikukawa:2002ms}, a general expression of the topological charge
was given under a certain restriction of gauge field configurations termed
$\U(1)$ embedding. In ref.~\cite{Kikukawa:2002ms}, it was also shown that a
single fundamental Weyl fermion in the chiral gauge reduced model gives rise to
an obstruction for a smooth fermion integration measure, if a Dirac operator
which obeys the Ginsparg-Wilson relation is adopted. The phenomenon is
analogous to the gauge anomaly in the original unreduced theory. It is then
natural to consider a cancellation of the obstruction among fermion multiplets
which belong to various gauge-group representations. As a first step toward
this consideration, in this work we study the topological charge for general
gauge-group representations in the vector-like reduced model. Our argument
below is applied to the quenched reduced model as well as the naive reduced
model \`a la Eguchi-Kawai. The extension to the twisted reduced model however
is not clear for the present.

\section{Axial anomaly with the GW relation}
The Ginsparg-Wilson (GW) relation
\begin{equation}
   \gamma_5D+D\gamma_5=D\gamma_5D
\end{equation}
ensures an exact chiral symmetry on the lattice and the corresponding axial
anomaly
\begin{equation}
   q(x)=\tr\gamma_5[1-\frac{1}{2}D(x,x)]
\label{two}
\end{equation}
has topological properties. For example, the topological charge~$Q$ defined by
\begin{equation}
   Q=\sum_xq(x)=\Tr\gamma_5(1-\frac{1}{2}D)=n_+-n_-
\label{three}
\end{equation}
becomes an integer identical to the index of the Dirac operator ($n_\pm$ is the
number of zero-modes of $D$ with $\pm$ chirality, respectively).

As~$D$, we adopt Neuberger's overlap Dirac operator
\begin{eqnarray}
   &&D=1-A(A^\dagger A)^{-1/2},\quad A=1-D_{\rm w},
\nonumber\\
   &&D_{\rm w}=\frac{1}{2}\sum_\mu[\gamma_\mu(\nabla_\mu^\ast+\nabla_\mu)
   +\nabla_\mu^\ast-\nabla_\mu],
\end{eqnarray}
where the covariant differences are defined by
\begin{eqnarray}
   &&\nabla_\mu\psi(x)=U_\mu(x)\psi(x+\hat\mu)-\psi(x),
\nonumber\\
   &&\nabla_\mu^\ast\psi(x)=\psi(x)-U_\mu^\dagger(x-\hat\mu)\psi(x-\hat\mu).
\end{eqnarray}
Since the topological charge~$Q$ is an integer, the overlap Dirac operator
becomes singular at certain points in the space of lattice gauge fields at
which the value of~$Q$ jumps. It is known that one can avoid these
singularities by requiring the gauge field to be ``admissible''
\begin{equation}
   \|1-U_\mu(x)U_\nu(x+\hat\mu)
   U_\mu(x+\hat\nu)^\dagger U_\nu(x)^\dagger\|<\epsilon,
\end{equation}
where $\epsilon$ is a positive constant smaller than $(2-\sqrt{2})/d(d-1)$.
Then the space of lattice gauge fields is divided into topological sectors and
the topological charge~$Q$ is well-defined for each topological sector.

When the gauge group is $\U(1)$, a complete parameterization of admissible
$\U(1)$ gauge fields has been known~\cite{Luscher:1999du}
\begin{equation}
   U_\mu(x)=\Lambda(x)V_\mu^{[m]}(x)U_\mu^{[w]}(x)
   e^{iA_\mu^{\rm T}(x)}\Lambda(x+\hat\mu)^{-1},
\label{seven}
\end{equation}
where $\Lambda(x)$ represents the gauge degrees of freedom and $V_\mu^{[m]}(x)$
has a constant field strength $F_{\mu\nu}(x)=2\pi m_{\mu\nu}/L^2$, where
$m_{\mu\nu}$ is the magnetic flux integer which labels topological sectors
($L$ is the size of the lattice). $U_\mu^{[w]}(x)$ carries the Wilson line,
$\prod_{s=0}^{L-1}U_\mu^{[w]}(s\hat\mu)=w_\mu\in U(1)$, and 
$A_\mu^{\rm T}(x)$ represents gauge invariant local fluctuations. With this
parameterization, the topological charge~$Q$ in eq.~(\ref{three}) is expressed
as~\cite{Igarashi:2002zz}
\begin{equation}
   Q=\frac{(-1)^{d/2}}{2^{d/2}(d/2)!}
   \epsilon_{\mu_1\nu_1\cdots\mu_{d/2}\nu_{d/2}}
   m_{\mu_1\nu_1}\cdots m_{\mu_{d/2}\nu_{d/2}}.
\label{eight}
\end{equation}

\section{Reduced model and the $\U(1)$ embedding}

The $\U(N)$ reduced model is defined by zero-volume limit of $\U(N)$ lattice
gauge theory: $U_\mu(x)\in\U(N)\to U_\mu\in\U(N)$. Here we consider the
fundamental fermion in the vector-like reduced model:
$\psi(x)\to\psi\in\hbox{fundamental rep.\ of $\U(N)$}$. The fermion sector
is defined by
\begin{equation}
   \langle\mathcal{O}\rangle_{\rm F}
   =\int\rmd\psi\rmd\overline\psi\,{\mathcal O}
   \exp(-\overline\psi D\psi).
\end{equation}
In this expression, the covariant difference in the overlap Dirac operator~$D$
is defined by $\nabla_\mu=U_\mu-1$. Then the admissibility condition and the
topological charge are given by
\begin{equation}
   \|1-U_\mu U_\nu U_\mu^\dagger U_\nu^\dagger\|<\epsilon,
   \quad Q=\tr\gamma_5(1-\frac{1}{2}D).
\label{ten}
\end{equation}
In ref.~\cite{Kiskis:2002gr}, an explicit example of admissible gauge field
configurations with~$Q\neq0$ was given. Here we use the following
method~\cite{Kikukawa:2002ms} which produces a wide class of admissible
configurations.

First we assume $N=L^d$ with an integer~$L$. Then the index of the fundamental
representation~$i$ ($1\leq i\leq N=L^d$) of~$\psi$ can be identified with a
site~$x$ on the lattice~$\Gammait$,
$\Gammait=\{\,x\in{\mathbb Z}^d\mid0\leq x_\mu<L\,\}$, by
$i(x)=1+x_d+Lx_{d-1}+\cdots+L^{d-1}x_1$ for example.

Next we assume that the reduced gauge field has the following form
(the $\U(1)$ embedding)
\begin{equation}
   U_\mu=u_\mu\Gammait_\mu,
\label{eleven}
\end{equation}
where~$u_\mu$ is an $N\times N$ {\it diagonal\/} matrix, and the ``shift''
matrix $\Gammait_\mu$ is defined by
\begin{equation}
   \Gammait_\mu={\mathbf 1}_L\otimes
   \cdots\otimes{\mathbf 1}_L\otimes U\otimes{\mathbf 1}_L
   \otimes\cdots\otimes{\mathbf 1}_L
\end{equation}
from $L\times L$ unit matrix ${\mathbf 1}_L$ and the $L\times L$ unitary
matrix~$U$
\begin{equation}
   U=\pmatrix{0&1&&\cr
              & \ddots & \ddots &\cr
              &        & \ddots &1\cr
              1&&&0\cr}
\end{equation}
which appears in the $\mu$-th entry. Then the covariant difference in the
reduced model assumes the form
\begin{equation}
  \nabla_\mu=U_\mu-1=u_\mu\Gammait_\mu-1.
\end{equation}
We note that all diagonal elements of~$u_\mu$ are pure-phase, i.e., they are
elements of~$\U(1)$. Since the matrix~$\Gammait_\mu$ realizes a shift on the
lattice, $\Gammait_\mu\psi_{i(x)}=\psi_{i(x+\hat\mu)}$, we see that the
fermion sector of the $\U(N)$ reduced model with the $\U(1)$
embedding~(\ref{eleven}) is completely identical to that of the $\U(1)$ lattice
gauge theory. In particular, the plaquette and the admissibility condition in
the reduced model are promoted to those of the $\U(1)$ lattice gauge theory. By
this way, we immediately find that the topological charge in the reduced model
in eq.~(\ref{ten}) is given by the expression~(\ref{eight}), where this time
integers~$m_{\mu\nu}$ parameterize a ``form'' of the admissible reduced
$\U(N)$ gauge fields~$U_\mu$~\cite{Kikukawa:2002ms}.

\section{$Q_R$ in the vector-like reduced model}

We next consider a fermion belonging to a general irreducible
representation~$R$ of~$\SU(N)$:
\begin{equation}
   (\psi)_{i_1,\ldots,i_n;j_1,\ldots,j_m},
\end{equation}
where each of $i_1$, \dots, $i_n$ transforms as the fundamental representation
and each of $j_1$, \dots, $j_m$ transforms as the anti-fundamental
representation. The covariant difference is then defined by
\begin{eqnarray*}
   &&(\nabla_\mu\psi)_{i_1,\ldots,i_n;j_1,\ldots,j_m}
\nonumber\\
   &&=(U_\mu)_{i_1k_1}\cdots(U_\mu)_{i_nk_n}
   (\psi)_{k_1,\cdots ,k_n;l_1,\cdots ,l_m}
\nonumber\\
   &&\quad\times(U_\mu^\dagger)_{l_1j_1}\cdots(U_\mu^\dagger)_{l_mj_m} 
   -(\psi)_{i_1,\ldots,i_n;j_1,\ldots,j_m}.
\end{eqnarray*}
For a such general representation, however, the $\U(1)$
embedding~(\ref{eleven}) itself does not provide a useful picture such as the
matrix-lattice correspondence which was utilized in
ref.~\cite{Kikukawa:2002ms}. Here we consider the large~$N$ approximation with
the $\U(1)$ embedding. We can set $\Lambda(x)=1$ and $A_\mu^{\rm T}(x)=0$ in
eq.~(\ref{seven}) because the topological charge~$Q$ is an integer-valued gauge
invariant quantity.

In terms of eigenvectors of the shift matrix
\begin{equation}
   \Gammait_\mu\phi(\vec p)
   =e^{2\pi ip_\mu/L}\phi(\vec p),
\end{equation}
where $\{\,\vec p\in{\mathbb Z}^d\mid0\leq p_\mu<L\,\}$, the topological charge
for the representation~$R$ is expressed as
\begin{eqnarray*}
   &&Q_R={1\over2}
   \sum_{\vec p_1}\cdots\sum_{\vec p_n}
   \sum_{\vec q_1}\cdots\sum_{\vec q_m}
\nonumber\\
   &&\quad\times
   \phi(\vec p_1)_{i_1}^\dagger\cdots\phi(\vec p_n)_{i_n}^\dagger
   \phi(\vec q_1)_{j_1}\cdots\phi(\vec q_m)_{j_m}
\nonumber\\
   &&\quad
   \times({H\over\sqrt{H^2}}{\cal P}_R)_{i_1,\ldots,i_n;j_1,
   \ldots,j_m;k_1,\ldots,k_n;l_1,\ldots,l_m}
\nonumber\\
   &&\quad\times\phi(\vec p_1)_{k_1}\cdots\phi(\vec p_n)_{k_n}
   \phi(\vec q_1)_{l_1}^\dagger\cdots\phi(\vec q_m)_{l_m}^\dagger.
\end{eqnarray*}
In this expression, $H=\gamma_5A$ and ${\cal P}_R$ is the projection operator
for the irreducible representation~$R$. It turns out that in the leading order
of the large~$N$ (or large~$L$; recall $N=L^d$) approximation, one can replace
the projection operator by
\begin{equation}
   {\cal P}_R\to{\dim R\over N^{n+m}}\times\hbox{identity operator}.
\end{equation}
We also note that, with the $\U(1)$ embedding, $u_\mu={\mathbf 1}_N(1+O(1/L))$
and $U_\mu=\Gammait_\mu(1+O(1/L))$ where sub-leading terms are diagonal
matrices and the operator~$H^2$ contains the commutator
$[U_\mu,U_\nu]=(e^{2\pi im_{\mu\nu}/L^2}-1)U_\nu U_\mu=
2\pi im_{\mu\nu}/L^2(1+O(1/L))$. With these observations and with the relations
$\sum_{\vec p_2}\cdots\sum_{\vec p_n}\sum_{\vec q_1}\cdots\sum_{\vec q_m}1=
N^{n+m-1}$ and
$\sum_{\vec p_1}=L^d\int_{-\pi}^\pi{\rmd^dk\over(2\pi)^d}(1+O(1/L))$
where $k_\mu=2\pi p_{1\mu}/L$, we see that the leading term in the large~$N$
approximation is given by the large~$L$ limit of
\begin{eqnarray}
   &&Q_R={\dim R\over N}{1\over2}L^d\int_{-\pi}^\pi{\rmd^dk\over(2\pi)^d}
\nonumber\\
   &&\quad\times\tr\gamma_5[-i\gamma_\mu s_\mu+1+\sum_\mu(c_\mu-1)]
\nonumber\\
   &&\quad\times\{
   s_\nu^2+[1+\sum_\nu(c_\nu-1)]^2
\nonumber\\
   &&\qquad
   -{2\pi i(n-m)m_{\nu\rho}\over 2L^2}\gamma_\nu\gamma_\rho c_\nu c_\rho
\nonumber\\
   &&\qquad
   +{2\pi i(n-m)m_{\nu\rho}\over L^2}\gamma_\nu ic_\nu s_\rho\}^{-1/2},
\label{nineteen}
\end{eqnarray}
where we have used the abbreviations $s_\mu=\sin k_\mu$ and~$c_\mu=\cos k_\mu$.
The integral~(\ref{nineteen}) is a well-known one in a calculation of the
{\it classical continuum limit\/} of the axial anomaly~(\ref{two}), if one
identifies $1/L$ with the lattice spacing~$a$ and $2\pi i(n-m)m_{\mu\nu}$ with
the field strength~$F_{\mu\nu}$ in the continuum theory. By this way, we have
\begin{equation}
   Q_R=(n-m)^{d/2}{\dim R\over N}Q_F(1+O(1/L)),
\end{equation}
where the topological charge for the fundamental representation~$Q_F$ is
given by eq.~(\ref{eight}).

The above calculational scheme will be very useful
to investigate a possible cancellation of the obstruction in chiral gauge
reduced models which was identified in ref.~\cite{Kikukawa:2002ms}. This study
is under progress.
For more details and for a more complete list of references, see
ref.~\cite{Inagaki:2003uu}.

\end{document}